\documentclass[fleqn,10pt]{wlscirep}

\usepackage{amsmath}
\usepackage{amssymb}
\usepackage{graphicx}
\usepackage{cite}

\usepackage{color}
\usepackage{multirow}
\usepackage{amsfonts}

\definecolor{red}{rgb}{1,0,0}

\title{Online Social Activity Reflects Economic Status}

\author[1,2]{Jin-Hu Liu}
\author[1,2]{Jun Wang}
\author[2,3,*]{Junming Shao}
\author[1,3,*]{Tao Zhou}
\affil[1]{CompleX Lab, Web Sciences Center, University of Electronic Science and Technology of China, Chengdu 611731, China.}
\affil[2]{Data Mining Lab, Web Sciences Center, University of Electronic Science and Technology of China, Chengdu 611731, China}
\affil[3]{Big Data Research Center, University of Electronic Science and Technology of China, Chengdu 611731, China.}
\affil[*]{E-mail: junmshao@uestc.edu.cn, zhutou@ustc.edu}

\begin{abstract}
To characterize economic development and diagnose the economic health condition, several popular indices such as gross domestic product (GDP), industrial structure and income growth are widely applied. However, computing these indices based on traditional economic census is usually costly and resources consuming, and more importantly, following a long time delay. In this paper, we analyzed nearly 200 million users' activities for four consecutive years in the largest social network (Sina Microblog) in China, aiming at exploring latent relationships between the online social activities and local economic status. Results indicate that online social activity has a strong correlation with local economic development and industrial structure, and more interestingly, allows revealing the macro-economic structure instantaneously with nearly no cost. Beyond, this work also provides a new venue to identify risky signal in local economic structure.
\end{abstract}

\begin{document}

\flushbottom
\maketitle
\thispagestyle{empty}

%---------------------------------------------------------------------------
%\section{Introduction}
%---------------------------------------------------------------------------
\section*{Introduction}
\label{sec:intro}
With the fall of the Iron Curtain and the upheaval of East European, the international environment has changed profoundly, in which security concerns are gradually dominated by economy instead of military considerations \cite{Anderson2000, Bell1976}. In fact, economic status not only drives the development of national defense, but more directly, affects our lives ranging from personal investment to policy-making in government \cite{Lewis2013, Rohe2013, Morgenstern2014}. For instance,  it is crucial for a mayor to consider the local economic structure in this city whenever he or she tends to make economic-related policies. As for ordinary people, the economic development usually accompanies with considerably improvement of the living standards. Meanwhile, our money-oriented plans or decisions explicitly or implicitly depend on the local or national economic status. How to provide an efficient and comprehensive view of the health of economy is thus of great importance for governments as well as individuals. Currently, economic census provides a straightforward way to present a clear picture of many facets of the national or global economy in terms of many indices such as Gross Domestic Product (GDP), industrial structure and income growth \cite{Sun2010, Balasubramanian2011, zhong2011}. However, computing these indices is usually a non-trivial task as they are always involved with considerable resources for a long time. For example, the widely accepted index, GDP, is defined as ``an aggregate measure of production equal to the sum of the gross values added of all resident institutional units engaged in production" \cite{OECD}. The calculation of such index needs to collect data from distinct local governments and then integrate all data together. During this process, there are two predominant factors arisen. Firstly, the economic census requires a large quantity of manual labour, materials and other related resources. And more importantly, the procedure is time consuming and thus many economy-oriented decisions cannot be made in time. Although some new economic census techniques have been developed in recent decades to speed up the process such as sampling, they often suffer in accuracy as the statistics are derived from the partial rather than the whole data. In light of these problems, in recent years a few indirect indices have been introduced to quickly reflect the economic status. A famous index, \emph{Keqiang Index} \cite{keqiang2007}, consisting of three indicators: the railway cargo volume, electricity consumption and loans disbursed by banks, is proposed to measure the economy of China. Although the index allows reflecting the economic development, it is insufficient to provide a comprehensive overview of the economy as the measure heavily relies on industry and largely ignores agriculture and services. The Producer Price Index (PPI) is another popular index, which is used to measure the average changes in prices received by domestic producers for their output \cite{VonDer2007}. This measure manifests the economic effects on people's daily life, providing a potential hint of the stability of economy and society. Although there also exist some more complicated indices such as Consumer Price Index (CPI), Social Retail Goods (SRG) and Foreign Direct Investment (FDI) to characterize economic development or economic structure,  it is a non-trivial task to obtain them due to their long-term  data collection and calculation procedures. Therefore, a fast, effective and comprehensive strategy to bring deep insight into the economic status is highly desired.

Social networks, such as Facebook, Twitter and Sina Microblog, are becoming the primary venues for people to obtain and share information on a global scale \cite{Khang2012,Bakshy2012}. With no doubt, the new social media is mainly driven by the advances of information technology \cite{Tran2013}. However, many researches also have demonstrated that national economy status and policies play an important role on the growth rate, diversity and stability of social networks \cite{Katona2008,Ioannides2007,Schweitzer2009,Shriver2013}. For instance, Katona \emph{et al.} \cite{Katona2008} have found that economy has significant effects on the structure of World Wide Web. Ioannides \emph{et al.}\cite{Ioannides2007} have studied individual outcomes in a dynamic environment in the presence of social interactions. Results show that the topology of social interactions is temporally changed once the individual outcomes vary continuously. Besides of the economists, socialists also point out that social networks and economic networks are mutually interrelated \cite{Jackson2010,eagle2010network,vscepanovic2015mobile}. Actually, social networks permeate our social and economic lives and play a central role in the transmission of information about job opportunities, and are critical to the trade of many goods and services\cite{Jackson2010}. Eagle \emph{et al.} \cite{eagle2010network} showed that the regional communication diversity is strongly correlated with economic development. Furthermore, the results in \cite{vscepanovic2015mobile} have showed high correlation in many of the cases revealing the diversity of socio-economic insights that can be inferred using only mobile phone call data. Moreover, Bettencourt \emph{et al.} \cite{Bettencourt2013} have shown the strong relationships among several social and economic indices. Beyond, some companies have attempted to utilize online networks properties to reflect economic status, e.g. Taobao CPI \cite{Liao2010}. Motivated by these studies, in this paper, we concentrate on analyzing and quantifying the latent relationship between economic status and online social activities, and propose a simple yet effective method to analyze the macro economic structure in a data mining framework. Building upon the data-driven analysis,  the work has several interesting findings: (i) The online social activity shows a strong linear correlation with economic development; (ii) The macro economic structure can be well reflected by social activity; (iii) Online social activity allows analyzing the economy status instantaneously and thus support in-time decisions from individuals to countries.

\section*{Materials and Methods}

\subsubsection*{Data Acquisition and Description}
 Here, to investigate the latent relationships between the social activities and economic status,  we focus on the primary social network in China: Sina Microblog (SM), and the economic data is derived from the National Statistic Bureau of the People's Republic of China.

\vspace{2mm}
\textbf{Sina Microblog.} This data was collected from Sina Microblog (www.weibo.com), which is the leading social network in China and was launched in 2009 by Sina Corporation. Like Twitter, approximate 100 million messages have been posted each day on this platform \cite{Sina20120228, Sina20130512}. Here we collect nearly 200 million online registered users from the year 2009 to 2012. For each microblogger, the registered date, verified information and location information including province and prefecture-level city are all examined. More than 97\% microbloggers are ordinary users (opposed to verified users). The basic statistics of the data set is summarized in Table \ref{tab:statistics}.

\begin{table*}[tb]
\centering \caption{\textbf{The statistics of user data of Sina Mircoblog.} $N$, $N_O$ and $N_L$ represent the total number of registered users, the ordinary users and the ordinary users with location information (including prefecture-level cities) from the year 2009 to 2012, respectively.}
\setlength\tabcolsep{20pt}
\begin{tabular}{ccccc}
\hline
User& 2009 & $2010$ & $2011$ & $2012$ \\
\hline
N& 2,025,595 & 38,263,550 & 82,934,212 & 75,242,922 \\
$N_O$& 1,841,346 & 35,370,466 & 79,462,793 & 73,718,770 \\
$N_L$ & 900,632 & 15,144,018 & 37,563,752 & 32,898,514 \\
\hline
\end{tabular}
\label{tab:statistics}
\end{table*}

\begin{figure}[!tb]
\begin{center}
\includegraphics[width=16cm]{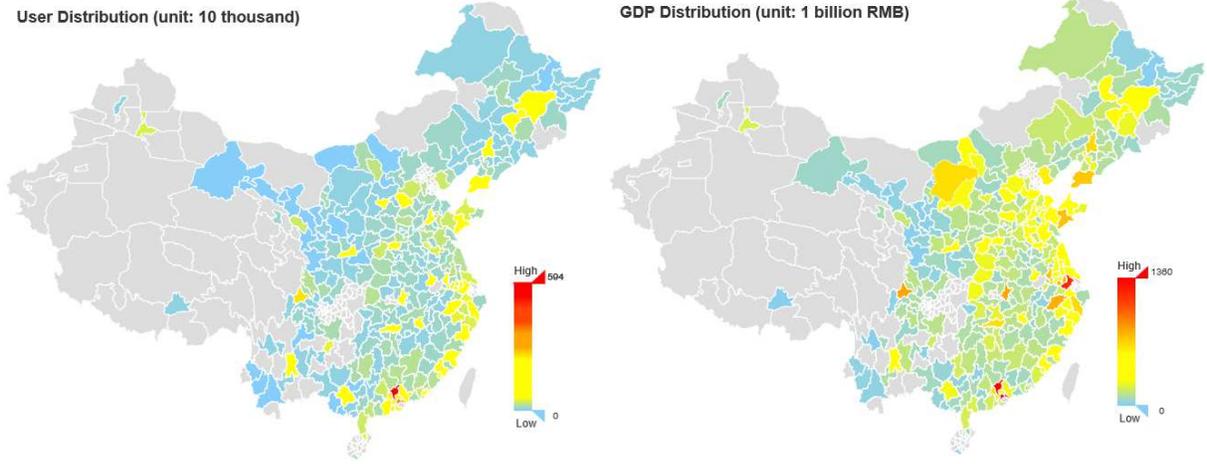}
\caption{The distributions of registered users in Sina Mircoblog (left) and the values of GDP (right) in the 282 prefecture-level cities of China in 2012.}
\label{Fig1}
\end{center}
\end{figure}

\vspace{2mm}
\textbf{National Economic Data}. The national economic data has been collected from the official book entitled ``China City Statistical Yearbook", which has been published by National Statistic Bureau (NSB) of the People's Republic of China in each year. In the statistical yearbook, major economic and social indices are reported, such as the total population, resident population, GDP, average GDP, industrial structure, to mention a few. Due to the time-consuming data collection and calculation procedures, the statistical yearbook cannot be published at the same year, but usually with about one year delay. From the books, we have collected and integrated the total population, resident population, GDP, average GDP and industrial structure of 282 prefecture-level cities (see Supplemental Information for the reason to choose prefecture-level cities as well as the list of city names.) in China  ranging from 2008 to 2012, respectively. The distributions of the number of registered users in Sina Mircoblog and the values of GDP in the 282 prefecture-level cities are presented in Figure \ref{Fig1}.

\subsubsection*{Correlation Analysis}
Here, we report two measures to exploit the relationship between online social activity and economic status upon the above two data sets.

\vspace{2mm}
\emph{Pearson correlation coefficient} is a measure to model the linear correlation between two random variables $X =\{x_1, \cdots,x_n\}$ and $Y=\{y_1, \cdots,y_n\}$ ($n$ is the dimension of $X$ and $Y$), yielding a value ranging from -1 (completely negative correlation) to 1 (completely positive correlation) \cite{Stigler1989} (see also the non-trivial bounds of Pearson correlation coefficient for heterogeneous systems \cite{guo2015memory}). Formally, the Pearson correlation coefficient $r$ is defined as:

\begin{equation}
r=\frac{\sum_{i=1}^n(x_i-\bar{x})(y_i-\bar{y})}{\sqrt{\sum_{i=1}^n(x_i-\bar{x})^2 \sum_{i=1}^n(y_i-\bar{y})^2}}.
\end{equation}
\label{def:pearson}
Here $\bar{x}=\frac{1}{N} \sum_{i=1}^n x_i$, $\bar{y}=\frac{1}{N}\sum_{i=1}^n y_i$ and $n$ is the dimension of $X$ and $Y$.

\vspace{2mm}
\emph{Spearman's rank correlation coefficient}  is a nonparametric measure of dependence between two variables. The strongest spearman correlation with value of 1 or -1 indicates a perfect monotone function of the other and there are no repeated values in the two variables \cite{Myers2010}. For calculation, both of variables, $X$ and $Y$, must be converted to ranks, $\overline{X}$ and $\overline{Y}$. And the $i$th element of $\overline{X}$, $\overline{x_i}$, represents ranking order of $x_i$ in all elements of $X$. Finally, Spearman's rank correlation coefficient $\rho$ is defined as follows.
\begin{equation}
\rho=1-\frac{6\sum{d_i^2}}{n(n^2-1)},
\label{def:spearman}
\end{equation}
where $d_i = \overline{x_i} - \overline{y_i}$ is the difference between ranks.

\subsubsection*{Predicting Economic Structure}
Here, we introduce support vector regression to uncover the macro economic structure based on online social activity.

\textbf{Support Vector Regression (SVR)}: The objective of SVR is to find a function $f(x)$ that has at most $\varepsilon$ deviation from the true target $y_i$ for all the training data, and is as flat as possible \cite{Cortes1995}.

\begin{equation}
f(x) = \langle w, x\rangle + b,  w \in \mathcal{X}, b \in \mathbb{R} ,
\label{def:svr}
\end{equation}

\begin{displaymath}
 \begin{array}{ll}
\textrm{minimize} & \frac{1}{2}\parallel w\parallel^2 , \\
\end{array}
\label{def:min}
\end{displaymath}
\begin{equation}
 \begin{array}{ll}
\textrm{subject to}&\left\{ \begin{array}{l}

y_i - \langle w, x\rangle - b\leq \varepsilon\\
\langle w, x\rangle + b - y_i\leq \varepsilon\\
\end{array}\right.
\textrm{.}
\end{array}
\label{def:min}
\end{equation}

To make SVR to handle nonlinear practical problems, the kernel trick is usually applied. The basic idea is to map the data into a high dimensional feature space $\mathcal{F}$ via a mapping function $\Phi$ and to do linear regression in the new space. Generally, there are several widely used kernel functions $K(x_i,x_j)$: (i) Linear kernel: $K(x_i, x_j)=(x_i \cdot x_j)$; (ii) Polynomial kernel: $K(x_i,x_j)=(x_i \cdot x_j+c)^{d}$, where $d>0$; (iii) Gaussian radial basis function (RBF): $K(x_i, x_j)=exp^{-\gamma||x_i-x_j||^2}$, where $\gamma>0$; (iv) Hyperbolic tangent: $K(x_i,x_j)=tanh(kx_i \cdot x_j+c)$. In this study, the Gaussian radial basis function is applied. In addition, to reduce variability of results and to evaluate how results generalize to an independent data set, we apply leave-one-out cross validation procedure \cite{Mosteller1948}.

\begin{figure*}[!tb]
\begin{center}
\includegraphics[width=15cm]{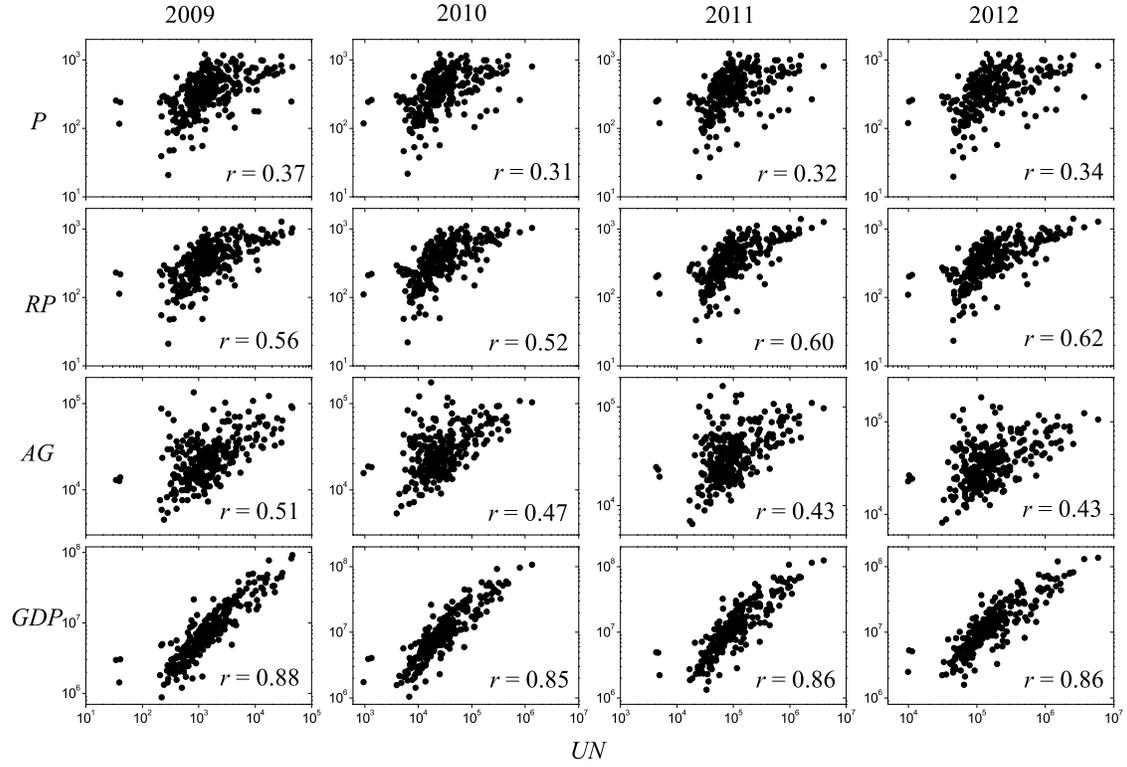}
\caption{\textbf{Relationships between the online social activity (i.e. the number of online registered users in Sina Microblog) and  economic indices from 2009 to 2012, respectively.} Here $P$, $RP$, $AG$, $GDP$ and $UN$ represent population, resident population, average gross domestic product, gross domestic product and the number of registered users in Sina Microblog, respectively.}
\label{Fig2}
\end{center}
\end{figure*}

\vspace{2mm}
\textbf{ Evaluation.} To evaluate the prediction performance, here we use two metrics: the \emph{root mean square error} (RMSE) and the \emph{relative error} (RE). For both metrics, the smaller the value is, the better the performance is.

\vspace{1mm}
RMSE is a popular metrics to capture the difference between the predicted value and true value. Formally, it is defined as:

\begin{equation}
RMSE=\sqrt{\frac{\sum_{i=1}^n{(\hat{y_i}-y_i)^2}}{n}},
\end{equation}
where $\hat{y_i}$ and $y_i$ are true and predicted values, respectively and $n$ is the size of the testing set. Given true value $\hat{y_i}$ and its predicted value $y_i$, if $\hat{y_i} \ne 0$, RE is defined as
\begin{equation}
RE =\frac{1}{n} \sum_{i=1}^n  \left| \frac{\hat{y_i} - y_i}{\hat{y_i}} \right|.
\end{equation}

\section*{Results}
\label{sec:results}

\begin{figure}[!t]
\begin{center}
\includegraphics[width=16cm]{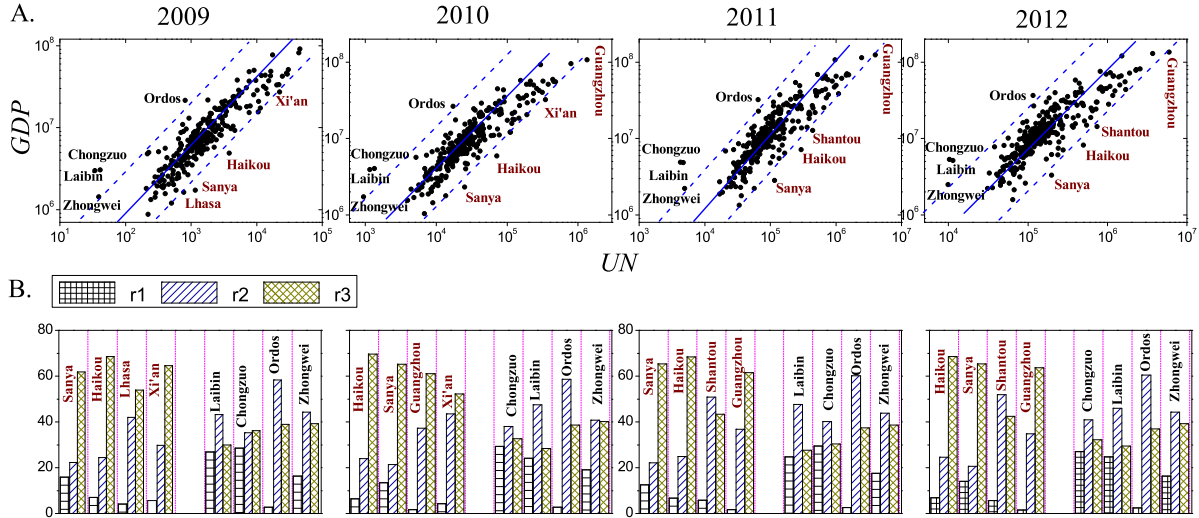}
\caption{\textbf{The correlation between online social activity and macro-economic structure.} (A) The scaling of the number of registered users with GDP during 2009 and 2012. The number on the top of each column represents the year. In each figure, the solid line is the fitting line via the least square method, and the dash line is parallel with the fitting line. (B) The histograms for the three-sector ratios of some selected outlier cities in distinct years as marked in (A). Here $r_1$, $r_2$ and $r_3$ represent the percentage proportions of primary, secondary and tertiary sectors, respectively.}
\label{Fig3}
\end{center}
\end{figure}

\subsubsection*{Relationship between Online Social Activity and Economic Status}
\label{sec:correlation}
Based on the correlation analysis, it is very interesting that the number of online registered users is highly related to economic indices such as population, resident population, average gross domestic product and gross domestic product (see Table \ref{tab:Correlation} and Figure \ref{Fig2}). Concretely speaking, the registered user number (UN) of Sina Microblog has positive correlations with population (P) ranging from 0.31 to 0.37 in different years. The result is in line with the real-world phenomenon that there will be more people use Sina Mircoblog if the local population itself is much larger. Similarly, UN has positive correlation with resident population (RP), but with a higher value. The reason is that RP accurately reflect the amount of inhabitants. In addition to population-related indices, UN also shows a strong correlation with the average gross domestic product (AG), suggesting that people in a richer city are more likely to use new social media. In contrast, the GDP has a surprisingly high correlation with UN (e.g., $r\approx 0.88$ and $\rho=0.90$), indicating that the city-level online social activity might be mainly determined by two factors: individual wealth and papulation size.

\begin{table*}[!htp]
\centering \caption{The correlation between UN and economic indices including P, RP, AG and GDP. $r$ and $\rho$ represent the Pearson correlation coefficient and spearman's rank correlation coefficient, respectively. In each row, the highest value is emphasized in bold.}
\setlength
\tabcolsep{8pt}
%\scriptsize
%\scriptsize
\begin{tabular}{c|cccc|cccc}
\hline
\multirow{2}{*}{} & \multicolumn{4}{c|}{Pearson Correction} & \multicolumn{4}{c}{ Spearman's Rank Correlation} \\
\cline{2-9}
& P-UN & RP-UN & AG-UN & GDP-UN & P-UN & RP-UN & AG-UN & GDP-UN \\
\hline
2009 & 0.3235 & 0.5574 & 0.5080 & \textbf{0.8780} & 0.5928 & 0.6836 & 0.5237 & \textbf{0.8988}  \\

2010 & 0.3117 & 0.5249 & 0.4661 & \textbf{0.8487} & 0.5995 & 0.7130 & 0.4906 & \textbf{0.8950}  \\

2011 & 0.3199 & 0.6025 & 0.4305 & \textbf{0.8560} & 0.5844 & 0.7130 & 0.4673 & \textbf{0.8709}  \\

2012 & 0.3390 & 0.6193 & 0.4269 & \textbf{0.8632} & 0.5944 & 0.7233 & 0.4427 & \textbf{0.8638}  \\

\hline
\end{tabular}
\label{tab:Correlation}
\end{table*}

%%%%%%%%%%%%%%% table 4 %%%%%%%%%%%%%%%%%
\begin{table}[!tbhp]
\centering \caption{The RMSE of the three methods to predict the distribution of GDP among the three sectors (primary, secondary and tertiary industries). In each row, the highest value is emphasized in bold.}
\setlength\tabcolsep{4pt}
%\scriptsize
\begin{tabular}{c|ccc|ccc|ccc}
\hline
\multirow{2}{*}{} & \multicolumn{3}{c|}{Primary} & \multicolumn{3}{c|}{Secondary} & \multicolumn{3}{c}{Tertiary} \\
\cline{2-10}
& Doo & EcoIndex & Random & Doo & EcoIndex & Random & Doo & EcoIndex & Random\\
\hline
2009 & 0.0846 & \textbf{0.0845} & 0.4603 & \textbf{0.1038} & 0.1080 & 0.3045 & \textbf{0.0800} & 0.0830 & 0.3378 \\
2010 & \textbf{0.0815} & 0.0817 & 0.4615 & \textbf{0.1016} & 0.1064 & 0.3087 & \textbf{0.0815} & 0.0850 & 0.3329 \\
2011 & \textbf{0.0805} & 0.0810 & 0.4587 & \textbf{0.0999} & 0.1042 & 0.3021 & \textbf{0.0830} & 0.0866 & 0.3337 \\
2012 & \textbf{0.0800} & 0.0807 & 0.4600 & \textbf{0.0985} & 0.1029 & 0.3002 & \textbf{0.0855} & 0.0881 & 0.3392 \\
\hline
\end{tabular}
\label{tab:RMSEStructure}
\end{table}

\subsubsection*{Uncovering Macro Economic Structure via Online Social Activity}
\label{sec:structure}
\begin{figure}[tb]
\begin{center}
\includegraphics[width=15cm]{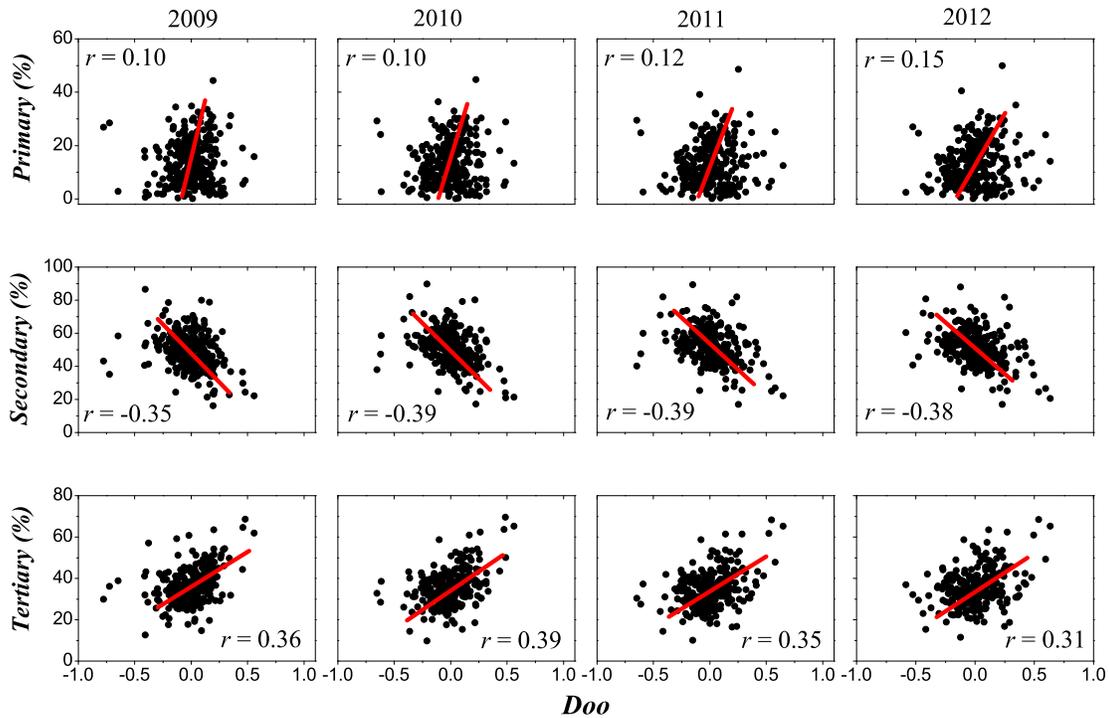}
\caption{\textbf{The relationships between \emph{Doo} and economic sectors.} $Doo$ represents the deviation between the fitting line and ture GDP in the logarithmic form. Primary (\%), Secondary (\%) and Teriary (\%) represent the percentage proportions of primary, secondary and tertiary sector to the whole GDP, respectively.  The relationships in different years, ranging from 2009 to 2012, are plotted as fifferent columns.}
\label{Fig4}
\end{center}
\end{figure}

As demonstrated above, the online social activity shows a strong relation with local economic status. Beyond that, the online social activity also well reflects the city-level macro economic structure. For each year, Figure \ref{Fig3}(A) marks the names of cities that far away from the fitting lines via the linear least square method. To bring deep insight into these cities, we found that the points below the fitting line (with larger number of registered users and a relatively lower GDP) tend to be service-driven cities. For instance, Lhasa, Sanya, Haikou and Xi'an, are identified as the top four cities with the largest distance below the fitting line in 2009. For all these cities, the common salient feature is the boom of tourism. In contrast, the cities over the fitting line focus on the heavy industry, such as Zhongwei, Laibin, Chongzuo and Ordos in 2009. In addition, these outliers are almost consistent across different years from 2009 to 2012. Figure \ref{Fig3}(B) further plots the macro economic structure (the distributions of the three sectors: primary, secondary and tertiary) of these cities in different years. Hence, to further test our hypothesis, we introduce a measure on the distance between the offline economic output (i.e. GDP) and online social activity (i.e. UN), with the fitting lines in Figure 3(A) as reference. Formally, the $Doo$ quantifies the deviation from the fitting line as:

\begin{equation}
Doo = l_i-y_i
\end{equation}
where $i$ is the label of the target city, $y_i=log_{10} GDP_i$ is the value of $i$'s GDP in the logarithmic coordinate and $l_i$ is the value on the fitting line for the registered number of users in the logarithmic coordinate at the city $i$. Figure \ref{Fig4} depicts the correlations between $Doo$ and the three sectors at different years. $Doo$ shows positive linear correlation with the primary and tertiary industry, the $Doo$ negative linear correlation with the secondary industry.

\begin{table*}[t]
\centering \caption{The performance of GDP prediction in terms of RMSE and RE using different information.  P-G: using the population of the last year to predict GDP of the next year. Similarly, RP, AG and Integrated stand for the cases of using the last year's resident population, average GDP and the integration of population, resident population and average GDP, respectively. UN-P: using the number of registered users to predict GDP in the same year. All variables above are expressed in logarithmic form. In each row, the highest value is emphasized in bold.}
\setlength\tabcolsep{6pt}
\begin{tabular}{ccccccccccccccc}
\hline
 & \multicolumn{2}{c}{P-G} & \multicolumn{2}{c}{RP-G}& \multicolumn{2}{c}{AG-G} & \multicolumn{2}{c}{Integrated-G} & \multicolumn{2}{c}{UN-G}\\ \cline{2-11}
 &RMSE &RE&RMSE &RE&RMSE &RE&RMSE &RE&RMSE &RE\\
\hline
2009 & 0.3212& 0.0358&	0.2980 &0.0332&	0.3161 &0.0365& 0.2916 &0.0338&	\textbf{0.1786}&\textbf{0.0203}\\
2010 & 0.3174& 0.0349&	0.2980 &0.0330&	0.3148 &0.0357& 0.2889 &0.0329& \textbf{0.1822}&\textbf{0.0203}\\
2011 & 0.3133& 0.0344&	0.2865 &0.0317&	0.3138 &0.0352& 0.2911 &0.0329&	\textbf{0.1952}&\textbf{0.0216}\\
2012 & 0.3062& 0.0332&	0.2728 &0.0299& 0.3147 &0.0351& 0.2952 &0.0331&	\textbf{0.1951 }&\textbf{0.0213}\\
\hline
\end{tabular}
\label{tab:GDPRMSE}
\end{table*}

Therefore, we expect to use the proposed index to predict the macro-economic structure. To objectively evaluate the prediction performance, three strategies were applied based on the support vector regression (see Materials and Methods). (a) $Doo$: using the single index $Doo$ to predict the distribution of GDP among three sectors; (b) $EcoIndex$: using the four traditional economic indices: P, RP, AG and GDP for prediction; (c) Random: randomly generate a value between the minimal value and the maximal value in the training set as the prediction. Table \ref{tab:RMSEStructure} gives a summary of the prediction performance of the three strategies. It is interesting to see that $Doo$ allows well reflecting the macro economic structure.

\subsubsection*{GDP Prediction}
\label{sec:prediction}
Considering the time and resources consuming procedure of GDP calculation in traditional economic census, we use UN to predict GDP building upon their latent strong correlation. Here we apply the highly related economic indices (population, residence population and average GDP) and online social activity (i.e. registered user number in Sina Microblog) to predict GDP, respectively. Table \ref{tab:GDPRMSE} gives the prediction accuracy based on different information, suggesting that the online social activity allows the most effective GDP prediction.

\section*{Discussion}
\label{sec:discusssion}
As the prosperity of economy is resulted from the aggregation of human activities, the economic status could be correlated with other human-related and human-activated systems, such as the structure of commercial webs \cite{Katona2008}, energy consumption \cite{Yang2000}, regional communication diversity \cite{eagle2010network} and city size \cite{Bettencourt2013}. Nowadays, online social networks, such as Facebook, Twitter and Sina Microblog, have been permeating every aspect of our social and economic lives. Therefore we need to give a close look at the relationship between online social activity and economic status. In this paper, to the first time, we quantitatively explore the potential relationship between the major economic indices of 282 prefecture-level cities of China and $2\times 10^8$ users' activities of the Chinese largest social networks (i.e., Sina Microblog). Empirical results show that the economic indices and the number of registered users in Sina Microblog is closely correlated (see Table \ref{tab:Correlation} and Figure \ref{Fig2}). Statistically speaking, people in more developed areas more frequently use the online social media.

The uncovered strong correlation further allows forecasting GDP in an effective and efficient way via support vector regression (see Table \ref{tab:GDPRMSE}). Comparing with the highly resource-consuming methods in traditional economic census, our methods have three remarkable properties. Firstly, due to the strong correlation between online social activity and economic status, it allows for accurate prediction. Secondly, the online social activity can be collected at any time, and thus we can analyze economy status instantaneously and support in-time decisions. Finally, collecting the online social activity has almost no cost compared to the national economic census.

More interestingly, the online social activity can reflect the macro-economic structure of cities and thus catch sight of outliers, some of which may have ill-posed economic structure and be fragile against the changes of the external economic environment. For instance, Ordos, located in Inner Mongolia, is one of the richest regions of China, whose nominal per-capita GDP is once ranked ahead of Beijing. This city has 1/6 of the national coal reserves, 1/3 of the national natural gas reserves and 1/2 of the national kaoline pockets. Apparently, coal mining, petrochemicals and production of building materials had become the pillars of its economy \cite{Ordos2010}. In its golden times, Ordos is not aware of the latent risk embedded in its less-diverse economic structure that is highly dependent on the prices of coal and gas. The housing price once rises to a unimaginable high level and recently, triggered by the drop of prices of coal and gas, the housing price of Ordos also drops very rapidly, leading to the overall economic collapse. Chongzuo, located in southwestern Guangxi, is also rich in minerals included manganese, gold, coal, and so on. It is the Chinese biggest manganese producer and the world's biggest producer of bentonite. In addition, its pillar industry is sugar refining \cite{Chongzuo2015}. With the decline sugar industry, like Ordos, Chongzuo is also exposed to adverse conditions through a process of economic restructuring \cite{Chongzuo201501}. Laibin \cite{laibin2012} and Zhongwei \cite{zhongwei2015} have undergone the similar tragedy for their over dependence on mineral price. All these four cities have been detected by our simple analysis (see Figure 3). On the other hand, economies of cities like Sanya and Haikou (see also Figure 3) heavily rely on the local services, which are very sensitive to the tourism. Although there are many previous studies showed the correlations with economic structure, such as environmental quality \cite{He2012}, warfare and political regime \cite{Karaman2013}, our study provides a simple, quick, yet effective way to predict the local macro-economic structure (see Table \ref{tab:RMSEStructure}), and more importantly, gives a hint to the health of local economic structure.

In summary, by coupling the population-level online social activity with economic outcomes in prefecture-level cities, we were able to conclude that the population-evolved activity of emerging online social media has a close relation with local economic status. Although causal relation cannot be established yet, online social activity seems to be a highly sensitive barometer of economic conditions. This interesting correlation further suggests a new venue for observing the health of economy, which may provide novel insights in addition to the statistics of the traditional economic census.

%---------------------------------------------------------------------------
%---------------------------------------------------------------------------


\begin{thebibliography}{0}

\bibitem{Anderson2000}
Anderson, E. W., Gutmanis, I. \& Anderson, L. D. {\em Economic Power in a Changing International System \/} (Psychology Press, London, 2000).

\bibitem{Bell1976}
Bell, D. The coming of the post-industrial society.
{\em Taylor \& Francis Group \/} {\bf 4}, 574--579 (1976).

\bibitem{Lewis2013}
Lewis, W. A. \& Anderson, L. D. {\em Theory of economic growth \/} (Routledge Press, NY, 2013).

\bibitem{Rohe2013}
Rohe, W. M., Van Zandt, S. \& McCarthy, G. The social benefits and costs of homeownership: a critical assessment of the research.
{\em The affordable housing reader \/} {\bf 40}, N.00--01 (2000).

\bibitem{Morgenstern2014}
Morgenstern, R. D. {\em Economic analyses at EPA: assessing regulatory impact \/} (Routledge Press, NY, 2014).

\bibitem{Sun2010}
Sun, Y. \& Du, D. Determinants of industrial innovation in China: Evidence from its recent economic census.
{\em Technovation \/} {\bf 30}, 540--550 (2010).

\bibitem{Balasubramanian2011}
Balasubramanian, N. \& Sivadasan, J. What happens when firms patent? New evidence from US economic census data.
{\em Rev. Econ. Stat. \/} {\bf 93}, 126--146 (2011).

\bibitem{zhong2011}
Zhong, W., Yuan, W., Li, S. X. \& Huang, Z.
The performance evaluation of regional R\&D investments in China: An application of DEA based on the first official China economic census data.
{\em Omega \/} {\bf 39}, 447--455 (2011).

\bibitem{OECD}
Organization for Economic Co-operation and Development. Gross Domestic Product (GDP).
(2001) Available at: http://stats.oecd.org/glossary/detail.asp?ID=1163. (Accessed: 10/08/2014).

\bibitem{keqiang2007}
The Economist. Keqiang Ker-ching: How China's Next Prime Minister Keeps Tabs on its Economy.
(2001) Available at: http://www.economist.com/node/17681868. (Accessed: 20/04/2015).

\bibitem{VonDer2007}
von der Lippe, P. {\em Index theory and price statistics \/} (Peter Lang Press, Oxford, 2007).

\bibitem{Khang2012}
Khang, H., Ki, E.-J., \& Ye, L. Social media research in advertising, communication, marketing, and public relations, 1997-2010.
{\em Journalism Mass Commun. Quart. \/} {\bf 89}, 279--298 (2012).

\bibitem{Bakshy2012}
Bakshy, E., Rosenn, I., Marlow, C., \& Adamic, L. The role of social networks in information diffusion.
in {\em Proc. of the 21st inter. conf. on World Wide Web (WWW' 12)\/} 519--528, Lyon, LYS, France (2012).

\bibitem{Tran2013}
Tran, S. T., Le, N. T., Nguyen, Q. B., Dang, B. P., Le, H. B. Introduction to information technology.
in {\em Proc. of the 9th inter. CDIO conf. (CDIO) \/} Cambridge, MA, USA (2013).

\bibitem{Katona2008}
Katona, Z. \& Sarvary, M. Network Formation and the Structure of the Commercial World Wide Web.
{\em Marketing Sci. \/} {\bf 27}, 764--778 (2008).

\bibitem{Ioannides2007}
Ioannides, Y. M. \& Soetevent, A. R. Social networking and individual outcomes beyond the mean field case.
{\em J. Econ. Behav. Organ. \/} {\bf 63}, 369--390 (2007).

\bibitem{Schweitzer2009}
Schweitzer, F., Fagiolo, G., Sornette, D., Vega-Redondo, F., Vespignani, A. \& White, D. R. Economic networks: The new challenges.
{\em Science\/} {\bf 325}, 422--425 (2009).

\bibitem{Shriver2013}
Shriver, S. K., Nair, H. S. \& Hofstetter, R. Social ties and user-generated content: Evidence from an online social network.
{\em Manage. Sci. \/} {\bf 59}, 1425--1443 (2013).

\bibitem{Jackson2010}
Jackson, M. O. {\em Social and economic networks \/} (Princeton University Press, Princeton, 2008).

\bibitem{eagle2010network}
Eagle, N., Macy, M. \& Claxton, R. Network diversity and economic development.
{\em Science \/} {\bf 328}, 1029--1031 (2010).

\bibitem{vscepanovic2015mobile}
{\v{S}}{\'c}epanovi{\'c}, S., Mishkovski, I., Hui, P., Nurminen, J. K., \& Yl{\"a}-J{\"a}{\"a}ski, A.
Mobile Phone Call Data as a Regional Socio-Economic Proxy Indicator.
{\em PLOS ONE \/} {\bf 10}, e0124160 (2015).

\bibitem{Bettencourt2013}
Bettencourt, L. M. A. The origins of scaling in cities.
{\em Science \/} {\bf 340}, 1438--1441 (2013).

\bibitem{Liao2010}
Liao, J., Wei, K. \& Chen, G. Relationship between Pricing and Customer's Perception C2C Commerce-Basing on Study of the Channel of C2C in Mainland China.
in {\em 2010 Inter. Conf. on E-Product E-Service and E-Entertainment (ICEEE) \/} Jiaozuo, Henan, CHN (2010).

\bibitem{Sina20120228}
Cao, B. Sina's Weibo Outlook Buoys Internet Stock Gains: China Overnight.
(2012) Available at:http://www.bloomberg.com. (Accessed: 20/08/2014).

\bibitem{Sina20130512}
Ong, J. China's Sina Weibo grew 73\% in 2012, passing 500 million registered accounts.
(2013) Available at:http://thenextweb.com. (Accessed: 20/08/2014).

\bibitem{Stigler1989}
Stigler, S. M. Francis Galton's account of the invention of correlation.
{\em Stat. Sci. \/} {\bf 2}, 73--79 (1989).

\bibitem{guo2015memory}
Guo, F., Yang, Z., Zhao, Z.-D. \& Zhou, T. Memory Constraints for Power-Law Series.
{\em arXiv: \/} 1506.09096 (2015).

\bibitem{Myers2010}
Myers, J. L., Well, A. \& Lorch, R. F. {\em Research design and statistical analysis \/} (Routledge Press, NY, 2010).

\bibitem{Cortes1995}
Cortes, C. \& Vapnik, V. Support-vector networks.
{\em Mach. Learn. \/} {\bf 20}, 273--297 (1995).

\bibitem{Mosteller1948}
Mosteller, F. A k-sample slippage test for an extreme population.
{\em Ann. Math. Stat. \/} {\bf 19}, 58--65 (1948).

\bibitem{Yang2000}
Yang, H. Y. A note on the causal relationship between energy and GDP in Taiwan.
{\em Energ. Econ. \/} {\bf 22}, 309--317 (2000).

\bibitem{Ordos2010}
Ordos Government. Natural resource (in Chinese).
(2010) Available at:http://www.ordos.gov.cn/zjordos/zrzy. (Accessed: 10/06/2015).

\bibitem{Chongzuo2015}
Chongzuo Government. The economic profile of Chongzuo (in Chinese).
(2015) Available at:http://www.chongzuo.gov.cn/front/newOnly?id=1026. (Accessed: 10/06/2015).

\bibitem{Chongzuo201501}
Robert. Chinese sugar industry get into a mess and urgently need transition (in Chinese).
(2014) Available at:http://finance.china.com.cn/roll/20140720/2551792.shtml. (Accessed: 10/06/2015).

\bibitem{laibin2012}
Yu, G. The profile of Laibin (in Chinese).
(2012) Available at:http://www.gx.xinhuanet.com/2012-03/22/c-111691189.htm (Accessed: 10/06/2015).

\bibitem{zhongwei2015}
Zhongwei Government. Mineral resources (in Chinese).
(2015) Available at:http://www.nxzw.gov.cn/zjzw1/kczy/584.htm (Accessed: 10/06/2015).

\bibitem{He2012}
He, J., Wang, H. Economic structure, development policy and environmental quality: an empirical analysis of environmental Kuznets curves with Chinese municipal data.{\em Ecol. Econ. \/} {\bf 76}, 49--59 (2012).

\bibitem{Karaman2013}
Karaman, K. K., Pamuk, S. Different Paths to the Modern State in Europe: The Interaction Between Warfare, Economic Structure, and Political Regime.
{\em Am. Pol. Sci. Rev. \/} {\bf 107}, 603--626 (2013).

\end{thebibliography}
\end{document}